\documentclass[12pt]{article}
\usepackage{amssymb,amsmath,amsthm,url,amscd,cite}
\usepackage{graphicx}
\usepackage{xcolor}
\usepackage{hyperref}
\usepackage[utf8]{inputenc}
\usepackage[T1]{fontenc}
\usepackage{times}
\newtheorem{theorem}{Theorem}

\newtheorem{lemma}[theorem]{Lemma}
\newtheorem{corollary}[theorem]{Corollary}

\newtheorem{proposition}[theorem]{Proposition}

\theoremstyle{definition}

\newtheorem{exe}{Example}

\newtheorem{definition}[theorem]{Definition}

\newcommand{\pf}{{\bf Proof. \ }}

\begin{document}
\title{ Some New Permutation Polynomials over Finite Fields}
\author{Nouara Zoubir and Kenza Guenda
\thanks{N. Zoubir and K. Guenda are with the Faculty of Mathematics, USTHB, Algiers, Algeria.}}
\maketitle

\begin{abstract}
%Permutation polynomials have numerous applications in mathematics and engineering.
In this paper, we construct a new class of complete permutation monomials and several classes of permutation polynomials. Further, by giving another characterization of o-polynomials, we obtain a class of permutation polynomials of the form $G(x)+ \gamma Tr(H(x))$, where G(X) is neither a permutation nor a linearized polynomial. This is an answer to the open problem 1 of Charpin and Kyureghyan
in [P. Charpin and G. Kyureghyan, When does $G(x)+ \gamma Tr(H(x))$ permute $\mathbb{F}_{p^n}$?, Finite Fields and Their Applications 15 (2009) 615--632].

\end{abstract}

\section{Introduction}

Let $\mathbb{F}_q$ be the finite field of order $q$.
A polynomial $P \in \mathbb{F}_q[x]$ is called a permutation polynomial over $\mathbb{F}_q $ if is a one to one map from $ \mathbb{F}_q$ to itself.
Permutation polynomials over a finite field have been a popular subject of study for many years
as they have numerous applications in areas such as coding theory, cryptography.
Details regarding properties, constructions and applications of permutation polynomials can be found in \cite{lidl} and \cite{handbook}.
Despite the tremendous interest, characterizing permutation polynomials and finding new families of permutation polynomials remain open questions.
Some recent progress on permutation polynomials can be found in \cite{akbary2, chapuy1, Fernando, lidl, handbook, Yuan}.

The remainder of this paper is organized as follows. In section \ref{Preli}, we give some preliminaries on permutation polynomials.
Section \ref{sec3} introduces  new classes of permutation polynomials; one class of monomial complete permutations over finite fields of odd characteristic, and two new classes of permutation polynomial from composed polynomials.
In section \ref{sec4} we construct several classes of permutation polynomials by using a combination from existing permutation polynomial and linearized polynomials. Further, we give another characterization of o-polynomials. This allows us to give an answer to the open problem 1 of Charpin and Kyureghyan \cite{kyureghyan1}. Namely, we give a class of permutation polynomial of the form $G(x)+ \gamma Tr(H(x))$, where $G(x)$ is neither a permutation nor linearized polynomial.

\section{Preliminaries}
\label{Preli}
In this section, some preliminary results are presented in order to be used in the sequel.
Let $p$ be a prime number and $q = p^n$, for a positive integer $n$ with a divisor $m \geqslant1$.
The trace function, denoted by $Tr_{m}^{n} (x)$ is a mapping from $\mathbb{F}_{q}$  to $\mathbb{F}_{p^m}$ defined as follow
\[
Tr_{m}^{n}(x)= x+x^{p^{m}} +x^{p^{2m}}+\ldots +x^{p^{n-m}}.
\]

The following result allows to characterize the permutation polynomial over a finite field.

\begin{lemma}\cite{lidl}
\label{lem:1.1}
The polynomial  $f \in \mathbb{F}_{q}$ is a permutation polynomial over $\mathbb{F}_{q}$ if and only if for every non zero $\gamma \in \mathbb{F}_{q}$,
\begin{equation}
\label{eq1}
\sum_{x \in \mathbb{F}_q}w^{Tr_{1}^{n}(\gamma f(x))}= 0,
\text{where } w \text{ is primitive p-th} \text{ root of unity}.
\end{equation}
\end{lemma}

When $f(x)$ is a monomial $x^n$ as a result of (\ref{eq1}) we obtain that:
\begin{equation}
\label{eq2}
x^{n} \text{ is a permutation polynomial of } \mathbb{F}_{q} \text{if and only if }\gcd(n,q-1)=1.
\end{equation}

\begin{lemma}\cite{Yuan}
\label{lem:2.2}
Let $L(x)$ be a linearized polynomial and let $Tr_{1}^{n}(x)$ be the trace function from $\mathbb{F}_{q}$ to $\mathbb{F}_p$.
Then for any $\alpha\in \mathbb{F}_{q}$ we have
\begin{equation}
\label{equ2}
L(Tr_{1}^{n}(\alpha))= Tr_{1}^{n}(L(\alpha)) = (\sum^{n-1}_{i=0} a_{i}) Tr_{1}^{n}(\alpha).
\end{equation}
\end{lemma}

A polynomial of the form
\[
L(x) = \sum^{n-1}_{i=0} a_{i}x^{p^{i}} \in \mathbb{F}_{p^{n}}[x],
\]
is called a linearized polynomial. A direct consequence of (\ref{eq1}) and (\ref{equ2}) is the fact that a linearized polynomial is a permutation polynomial over $\mathbb{F}_{q}$ if and only if $0$ is the only root of $L(x)$ in $\mathbb{F}_{q}$.

The following results will be useful later.

\begin{lemma}\cite[Theorem 3.75]{lidl}
\label{lem3}
 Let $r\geq2$ be an integer and $a$ in $\mathbb{F}_{q}^{*}$ then the binomials $x^{r} - a$ is irreducible in $\mathbb{F}_{q}[x]$ if and only if the following two conditions are satisfied:\\
 \begin{enumerate}
\item[(i)] each prime factor of $r$ divide the order $e$ of  $a$  in $\mathbb{F}_{q}^{*}$; but not $(q-1)\diagup e$;\\
\item[(ii)] $q \equiv 1\mod 4$  if $ r \equiv 0 \mod 4.$
\end{enumerate}
\end{lemma}

\section { New Classes of Permutation Polynomials}
\label{sec3}
In this section we give new classes of permutation polynomials.

\subsection{Complete Permutation Monomials}
In this section we give a class of complete permutation monomials.
Further, these results are extended to obtain other classes of permutation polynomials.
We begin with the following definition.
\begin{definition}
A polynomial $f(x)\in \mathbb{F}_{q}[x]$ is called a complete permutation polynomial (CPP) if and only if $f(x)$ and
$f(x) + x$ are both permutation polynomials over $\mathbb{F}_{q}$.
\end{definition}
The following lemma will be used later.
\begin{lemma}\cite{Jacobson}
\label{Jacobson}
The polynomial $x^2+1$ has a solution in $\mathbb{F}_{p^m}$ if and only if $p\equiv 1 \mod4 $ or $ p \equiv 3 \mod 4$ and $m$  is even.
\end{lemma}

From Lemma \ref{Jacobson}, if $m$ and $p$ are such that $p^m\equiv 3 \mod 4$,
the polynomial $x^2+1$ is irreducible over $\mathbb{F}_{p^m}$. Let $\alpha$ be a root of such polynomial in an extension field.
Then $\alpha$ has an order $4$ in the multiplicative group of $\mathbb{F}_{p^{2m}}$ = $\mathbb{F}_{p^m}(\alpha)$
and $\alpha^{p^m} = \alpha^3$, so it can be concluded that
\begin{equation}
\label{eq:2,2}
 Tr_m ^ {2m}(\alpha) = Tr_m ^ {2m}(\alpha^3) = 0, Tr_m ^ {2m}(\alpha^2) = -2.
\end{equation}
Now, let $x$  be an arbitrary element of $\mathbb{F}_{p^{2m}}$, then $x = x_0 +x_1\alpha,$    $x_0, x_1 \in\mathbb{F}_{p^m}.$
Since the trace function is additive, we have then
\begin{equation}
\label{eq:3}
 Tr_m ^ {2m}(x) = Tr_m ^ {2m}(x_0 +x_1\alpha) = 2x_0.
\end{equation}

Now we give our first result which is a generalization of \cite[Theorem 3.3]{xu}.
\begin{theorem}
\label{th8}
Let $p$ be a prime integer and $m$ be an odd positive integer such that $p^m \equiv 3 \mod 4$. Further, assume that $\gcd(p^{2m}-1, p^m + 2)=1$ and $\vartheta$ is a nonzero element in
$\mathbb{F}_{p^{2m}}$ with $Tr_m ^ {2m}(\vartheta) = 0$. Then the monomial $\vartheta^{-1} x^{p^m + 2} $ is a CPP over $\mathbb{F}_{p^{2m}}$.
\end{theorem}
\pf
Denote
\begin{equation}
 S = \{v_0 + v_1\alpha: v_0, v_1 \in\mathbb{F}_{p^m},v_0 = 0\} \backslash\{0\},
\end{equation}
and let $\alpha$ be a root of $x^2+1$ in $\mathbb{F}_{p^{2m}}$.
From (\ref{eq:2,2}) and (\ref{eq:3}), $S$ is the set of nonzero elements $\vartheta$ in $\mathbb{F}_{p^{2m}}$ with $Tr_m ^ {2m}(\vartheta) = 0$.
For each $\vartheta \in S$, then the assumption  $\gcd(p^{2m}-1, p^m + 2) = 1$ gives from (\ref{eq2}) that,
the monomial $\vartheta^{-1} x^{p^m + 2}$ is a permutation polynomial over $\mathbb{F}_{p^{2m}}$.
To prove that $\vartheta^{-1} x^{p^m + 2} $ is a CPP over $\mathbb{F}_{p^{2m}}$, it is sufficient to show that $x^{p^m + 2}+ \vartheta x$
is a permutation polynomial over $\mathbb{F}_{p^{2m}}$ for each $\vartheta \in S$.
As $\gcd (p^{2m}-1, p^m + 2)= 1$, hereafter a nonzero $\gamma\in\mathbb{F}_{p^{2m}}$ will be represented as
$\gamma = \beta^{p^m + 2}$ for a unique nonzero $\beta \in\mathbb{F}_{p^{2m}}$.
Then we have
\[
\begin{array}{ccl}
\sum_{x \in \mathbb{F}_{p^{2m}}} w^{{Tr_1 ^ {2m}}(\gamma(x^{p^m+2} + \vartheta x))}
&=&\sum_{x \in \mathbb{F}_{p^{2m}}} w^{{Tr_1 ^ {2m}}((\beta x)^{p^m+2} + \beta ^{p^m+1} \vartheta (\beta x))}\\
&=&\sum_{x \in \mathbb{F}_{p^{2m}}} w^{{Tr_1 ^ {2m}}(x^{p^m+2} + \beta ^{p^m+1} \vartheta x)}\\
&=& \sum_{x \in \mathbb{F}_{p^{2m}}} w^{{Tr_1 ^ m}(Tr_m ^ {2m}(x^{p^m+2} + \beta ^{p^m+1} \vartheta x))}.
\end{array}
\]
Expressing $x \in\mathbb{F}_{p^{2m}}$ as $ x _0 + x _1\alpha$, from (\ref{eq:2,2}) we have
\[
\begin{array}{ccl}
Tr_m ^ {2m}( x^{p^m+2}) = Tr_m ^ {2m}((x _0 + x _1\alpha)^{p^m+2})
&=& Tr_m ^ {2m}((x _0 + x _1\alpha)^{p^m}(x _0 + x _1\alpha)^2)\\
&=& Tr_m ^ {2m}(x _0^3 + 2 x _0 x _1^2 +( 2 x _0^2 x _1 + x _1^3)\alpha + x _0 x _1^2 \alpha^2 +x _0^2 x _1 \alpha ^3)\\
&=& 2(x _0^3 + 2 x _0 x _1^2),\\
\end{array}
\]
since $m$ is odd then, $\alpha^{p^m} = \alpha^3$.

As $(\beta^{p^m +1})^{p^m-1} = 1$, we have that $\beta^{p^m +1}\in \mathbb{F}_{p^m} $ and $\beta^{p^m +1}\vartheta \in S$.
From (\ref{eq:3}) we can assume that $\beta^{p^m +1}\vartheta  = u = u_1\alpha$ with $u_1 \in\mathbb{F}_{p^m }$ and then
\[
\begin{array}{ccl}
Tr_m ^ {2m}(\beta^{p^m +1}\vartheta x) &=& Tr_m ^ {2m}(u_1\alpha(x _0 + x _1\alpha))\\
&=& Tr_m ^ {2m} (u_1\alpha x_0 + u_1 x_1{\alpha}^2)\\
&=& u_1 x_1.
\end{array}
\]

Combining this, with the fact that $Tr_1 ^ {2m}( z^{p^m+2}) = Tr_1 ^ {2m}( z)$ for any $z \in \mathbb{F}_{p^m }$, we have

\begin{equation}
\label{equ6}
\begin{array}{ccl}
\sum_{x \in \mathbb{F}_{p^{2m}}} w^{{Tr_1 ^ {2m}}(\gamma(x^{p^m+2} + \vartheta x))}
&=& \sum_{x \in \mathbb{F}_{p^{2m}}} w^{{Tr_1 ^ m}(Tr_m ^ {2m}(x^{p^m+2} + \beta ^{p^m+1} \vartheta x))}\\
&=& \sum_{{x_0}, { x_1} \in \mathbb{F}_{p^{m}}} w^{{Tr_1 ^ m}(2x _0^3 ( x _1^6 + 1) + u_1 x_1)}\\
&=& \sum_{ {x_1} \in \mathbb{F}_{p^{m}}} w^{{Tr_1 ^ m}(u_1 x_1)}\sum_{{x_0} \in \mathbb{F}_{p^{m}}} w^{{Tr_1 ^ m}(2x _0^3 ( x _1^6 + 1))} \\
&=&0.
\end{array}
\end{equation}
Since the equation $x _1^6 + 1 = 0$ has no solution in $\mathbb{F}_{p^{m}}.$\\
Hence, for every nonzero $\gamma \in \mathbb{F}_{p^{2m}}$, we have
\[
\sum_{x \in \mathbb{F}_{p^{2m}}} w^{{Tr_1 ^ {2m}}(\gamma(x^{p^m+2} + \vartheta x))} = 0
\]

\qed
\begin{exe}
\label{exe3}
If $\omega$ generates $\mathbb{F}_9^* $ such that $\omega^2+1 = 0$  and $Tr_{1}^{2} ( \omega) = 0$,
then the monomial $\omega^{-1} x^5$ is a CPP  over $\mathbb{F}_9$.
\end{exe}

The following result  allows  to construct a new class of complete permutation polynomials.
\begin{lemma}\cite[Theorem 2]{Niedereter}
\label{lemm1}
 If f(x) is a CPP of $\mathbb{F}_q$, then  the  polynomial $f(x + a) + b$ is also a CPP, for all $a, b \in \mathbb{F}_q$.
\end{lemma}
%We use Lemma~\ref{lemm1} and we apply the first condition for a monomial $x^{p^m+2}$. We obtain the following Corollary.

\begin{corollary}
Under the assumptions of Theorem~\ref{th8}; the polynomial $f(x) = \vartheta^{-1} x^{p^m + 2} $ being a CPP over $\mathbb{F}_{p^{2m}}$, then
 \[
\begin{array}{ccl}
 f(x + a) + b = \vartheta^{-1}(x^{p^m+2} + a^{p^m}x^2 + 2ax^{p^m+1} + 2 a^{p^m+1}x + a^2x^{p^m} + a^{p^m+2}) + b\\
\end{array}
\]
permute $\mathbb{F}_{p^{2m}}$, for all $a$ and $b \in \mathbb{F}_{p^{2m}}$.
\end{corollary}

\pf
The proof is easily obtained by applying Theorem~\ref{th8} and Lemma~\ref{lemm1}.
\qed

\begin{exe}
The polynomial $f(x) = \vartheta^{-1} x^{29} $ be a CPP over $\mathbb{F}_{729}$. Then
$ f(x + a) + b = \vartheta^{-1}(x^{29} + a^{27}x^2 + 2ax^{28} + 2 a^{28}x + a^2x^{27} + a^{29}) + b$ is a CPP over $\mathbb{F}_{729}$.
\end{exe}

\subsection{Permutation Polynomials from Composed Polynomials}

Wan and Lidl \cite{Wan} gave a characterization of permutation polynomials using composed polynomials.
Since then, several classes of permutation polynomial have been obtained using the characterization of Lidl and Wan.
Recently, Laigle-Chapuy \cite{chapuy1} gave the the following modified version of the result in \cite{Wan}.
\begin{proposition}\cite[Theorem 3.1]{chapuy1}
\label{lem11}
Let $p$ be a prime, $s$ be a positive integer and $k$ be the order of $p$ in $\mathbb{Z}/s\mathbb{Z}$.
Let $l$ be a positive integer and $q = p^{kls}$.
Assume $r$ is a positive integer coprime with $q -1$ and $P(x)$ is a polynomial in $\mathbb{F}_{p^{kl}}[x]$.
Then the polynomial $x^rP(x^\frac{q-1}{s})$ is a permutation polynomial over $\mathbb{F}_{q}$ if and only if
\[
\forall\omega \in \mathbb{F}_{q} \text{ such that }\omega^s = 1, P(\omega)\neq 0.
\]
\end{proposition}

Using Proposition~\ref{lem11} and our previous class of CPP we construct a new class of permutation polynomial.

\begin{theorem}
\label{theo25}
Let $ P(x) = \vartheta^{-1} x^{p^m + 2} $ be the complete permutation polynomial over $\mathbb{F}_{p^{2m}}$ given in Theorem \ref{th8} and $\omega$ is an s-th root of unity with $s\mid {p^{2m}-1}$. If $gcd(r, p^{2ms}-1) = 1$ and $\vartheta$ is a nonzero element in
$\mathbb{F}_{p^{2m}}$ with $Tr_m ^ {2m}(\vartheta) = 0$, then $x^rP(x^\frac{p^{2ms}-1}{s})$ is a permutation polynomial over $\mathbb{F}_{p^{2ms}}$.
\end{theorem}
\pf
Assume that $P(x) = \vartheta^{-1} x^{p^m + 2}$ is a complete permutation polynomial over $\mathbb{F}_{p^{2m}}$ and
$\omega$ is an s-th root of unity with $s\mid{p^{2m}-1}$.
Assume that $gcd(s, p^m + 2)= d$, under our hypotheses we have to prove that $ P(\omega) \neq 0$.
From Bezout theorem there exist two integers $u$ and $v$ such that $s.u +  (p^m + 2)v = d$.
If $P(\omega) = \vartheta^{-1}\omega^{p^m + 2} = 0$ then $ \omega^d = \omega^{({p^m + 2})v + su} =  0$ for $d\mid s$. Thus this is a contradiction to the fact that $\omega$ is an $s-th$ root of unity.
Then from Proposition~\ref{lem11} the polynomial $x^rP(x^\frac{p^{2ms}-1}{s})$ is a permutation polynomial in $\mathbb{F}_{p^{2ms}}$.
On the other hand assume that  $Q(x) = \vartheta^{-1} x^{p^m + 2} + x$ is also a permutation polynomial and
$\omega$ is an $s-th$ root of unity with $s\mid{p^{2m}-1}$. Then if $Q(\omega) = \vartheta^{-1}\omega^{p^m + 2} + \omega = 0$, this gives $\omega^{p^m + 1} = - \vartheta$.
Since from the hypotheses of Theorem~\ref{th8} we have  $Tr_m ^ {2m}(\vartheta) = 0$ which is equivalent to $\vartheta^{p^m} + \vartheta = 0.$
 Hence $ Q(\omega)  = \vartheta^{-1} \omega^{p^m + 2} + \omega = 0$, becomes $\omega^{p^m + 1} = -\vartheta$, which is equivalent to $-2\omega^{p^m + 1} = 0$. Thus, this is a contradiction to the fact that $\omega$ is an $s-th$ root of unity.
 Then from Proposition~\ref{lem11} the polynomial $x^rQ(x^\frac{p^{2ms}-1}{s})$ is a permutation polynomial in $\mathbb{F}_{p^{2ms}}$.
\qed

\section{Permutation Polynomials from Linearized Polynomials}
\label{sec4}
In this section, using a combination of existing permutation polynomials and linearized polynomials, new permutation polynomials are constructed.
\begin{proposition}\cite[Theorem 1]{Marcos}
\label{prop:9}
Let $g(x)\in \mathbb{F}_q [x]$ be a permutation polynomial and $L(x)\in \mathbb{F}_{q} [x]$ be a linearized polynomial. If $g(x) + L(x)$ is also a permutation polynomial over $\mathbb{F}_{q}$,
then $f(x) = g^{-1}(L(x) + \delta) + x$ permutes ${\mathbb{F}_q}$, for any $\delta \in \mathbb{F}_q$.
\end{proposition}
Next we generalize Proposition \ref{prop:9}.

\begin{proposition}
\label{prop:10}
Let $\mathbb{F}_q [x]$ be the finite field of order $q$ and characteristic $p$, $g(x)\in \mathbb{F}_q [x]$ be a permutation polynomial and $L(x)\in \mathbb{F}_{q} [x]$ be a linearized polynomial.
If $g(x) + L(x)$ is also a permutation polynomial over $\mathbb{F}_{q}$,
then $f(x) = g^{-1}(L(x^p) + \delta) + x^p$ permutes ${\mathbb{F}_q}$, for any $\delta \in \mathbb{F}_q$.
\end{proposition}
\pf
Let $c$ be an element in $\mathbb{F}_{q}$, then there exists an $a\in \mathbb{F}_{q}$ such that $c = a^{p}$.
Consider
\begin{equation}
\label{eq:1}
g^{-1}(L(x^{p}) + \delta)+ x^{p} = c,
\end{equation}
which is equivalent to
\[
L(x^{p}) + \delta = g((a- x)^{p}),
\]
and let $ y = (a- x)^{p}$, then $L(c) + \delta = g(y) + L(y)$ is a permutation polynomial, so there exists a unique $y$ satisfying this equality.
Thus, there exists a unique $x \in \mathbb{F}_q $ satisfying (\ref{eq:1}).
\qed

\begin{exe}
\label{exe8}
Let $L(x) = x^3+x$ be linearized polynomial over $\mathbb{F}_{3^2}$ and let $g(x)= -\omega x^5$  be a CPP.
Then the polynomial \[g^{-1}(L(x^3)) + x^3  = \omega^5( x^3 +x) + x^3\]
permutes $\mathbb{F}_{3^2}$.
\end{exe}

\begin{exe}
From \cite{aml} the polynomial $g(x)= 2x^{7}+3x \in \mathbb{F}_{7^{2}}[x]$ is a permutation polynomial
 and its inverse is  $g^{-1}(x) = x^{7}+4x$. Now, let $L(x)= x\in \mathbb{F}_{7^{2}}[x]$ be a linearized polynomial over $\mathbb{F}_{7^{2}}[x]$
so then $g(x)+L(x) = 2x^{7}+4x $ is a permutation polynomial according to Proposition~\ref{prop:10}.
Then $g^{-1}(L(x^{7}))+ x^{7} = 5x^{7}+x$ permutes $\mathbb{F}_{7^{2}}$.
\end{exe}

\begin{exe}
For $p \equiv 1 \mod 3$, let $g(x)= x^{p^2}\in \mathbb{F}_{p^{3}}[x]$
be a CPP and $g^{-1}(x) = x^{p}$ its inverse modulo $x^{p^{3}}-x$.
Further, let $L(x)= x$ be a linearized polynomial over $\mathbb{F}_{p^{3}}[x]$,
so then $g^{-1}(L(x^{p}))+ x^{p} = x^{p^2} + x^{p}$ permutes $\mathbb{F}_{p^{3}}$.
\end{exe}

For $g(x)$ be a CPP, if we take $L(x)= x$ in Proposition~\ref{prop:10} then we obtain the following Corollary.

\begin{corollary}
\label{coro14}
Let $s$ and $t$ be two positive integers, such that $g(x) = x^s$  and $g^{-1} = x^t$  are CCP. Then for  linearized polynomial $L(x)$ and $\delta $ in $\mathbb{F}_{q} $ the polynomial $ (x^p + \delta)^t + x^p$ permute $\mathbb{F}_{q} $,
\end{corollary}

In \cite{akbary2}, Akbary, Ghioca and Wang proposed the following lemma.

\begin{lemma}
\label{lem10}
Let $A$, $S$ and $\bar{S}$ be finite sets with
$\sharp{\bar{S}}$ = $\sharp S$ and let
$f: A \rightarrow A$,
$h: S \rightarrow \bar{S}$, $\lambda: A \rightarrow S$ and
$\bar{\lambda}: A \rightarrow \bar{S}$
be maps such that $\bar{\lambda}of$ = $h o\lambda$.
If both $\lambda$ and $\bar{\lambda}$ are surjective, then the following statements are equivalent:
\begin{enumerate}
\item[(i)]$f$ is a bijection (a permutation over $A$); and
\item[(ii)]$h$ is a bijection from $S$ to $\bar{S}$ and $f$ is injective on $ \lambda^{-1}(s)$ for each $s \in S$.
\end{enumerate}
\end{lemma}

Next, we use the result given by Yuan and Ding in \cite{Yuan} and Lemma~\ref{lem10} to obtain a new class of permutation polynomials of the type $f(x) = u(x) + v(x)$ given in the next Theorem.

\begin{theorem}
\label{the15}
Let $L_1(x)$ and $L(x)$ be two linearized polynomials over $\mathbb{F}_{p^2} $ with $p > 2$ such that $L(Tr_{1}^{2}(L_1(x))) = 0$.
Then the polynomial $f(x)= L_1(x) +Tr_{1}^{2}(L_1(x))$ permutes $\mathbb{F}_{p^2} $ if and only if $L_1(x)$ permutes $\mathbb{F}_{p^2}$.
\end{theorem}
\pf
Define $S = \{ x^p + x^{p-1} +\delta, x \in\mathbb{F}_{p^2} \}$ for $\delta \in\mathbb{F}_{p^2}$ and $\sharp\bar{S} = \sharp S = p^2 - p$,
and $ \bar{S} =  \{ x^p + x^{p-1} , x \in\mathbb{F}_{p^2 } \}$
we define $\lambda(x) = \bar{\lambda}(x) = L(x)$ is a linearized polynomial such that $\lambda$ is a surjection from $\mathbb{F}_{p^2}$ to $S$
and $\bar{\lambda}$ is a surjection from $\mathbb{F}_{p^2}$ to $\bar{S}$ and is additive and $h(x) = L_1(x)$ is a function from $S$ to $\bar{S}$.
Define $u(x) = L_1(x)$ and $v(x) = Tr_{1}^{2}(L_1(x))$ for $x \in\mathbb{F}_{p^2}$.
Then $u(x)$ and $v(x)$ are functions from $\mathbb{F}_{p^2}$ to $\mathbb{F}_{p^2}$.

It follows from Lemma~\ref{lem10} and the assumptions that $u(x)+v(x)$ permutes $\mathbb{F}_{p^2}$ if and only if $h$ is a bijection from $S$ to $\bar{S}$
and $u(x)+v(x)$ is injective on each $\lambda^{-1}(s)$ for all $s \in S$.
On the other hand, we have that $L(Tr_{1}^{2}(L_1(x)))= 0$ for every $x\in\mathbb{F}_{p^2}$ and $v(x) = Tr_{1}^{2} (L_1(x))$ is a constant on each ${\lambda}^{-1}(s)$ for all $s\in S$.
Hence $\bar{\lambda}(u(x)+v(x))$ = $\bar{\lambda}(u(x))+\bar{\lambda}(v(x)) = \bar{\lambda}(u(x))= h(\lambda(x))$
for all $x\in\mathbb{F}_{p^2 }$.
Then $h : S\rightarrow \bar{S}$ is surjective and thus bijective because $S$ and $\bar{S}$ are finite sets of the same cardinality.
It then follows that $u(x)+v(x)$ is injective on each $\lambda^{-1}(s)$ for all $s \in S$ if and only if $u(x)$ is injective on each $\lambda^{-1}(s)$ for all $s \in S$.
Hence all of the conditions in Lemma~\ref{lem10} are satisfied.
Then the map $f(x) = u(x) + v(x)$  permutes $\mathbb{F}_{p^2}$ if and only if $u(x)$ permutes $\mathbb{F}_{p^2}$.
\qed

\begin{corollary}
\label{coro17}
Let $ p \equiv 1 \mod 4$, let $L_1(x)$  and  $L(x)$ be linearized polynomials over $\mathbb{F}_{p^2}$, such that $L_1(x)$ is a permutation polynomial and $L(Tr_{1}^{2}(L_1(x))) =0$ .
Then the polynomial
\[
f(x) = L_1^{-1}( Tr_{1}^{2}(L_1(x))+\delta)+x^p,
\]
permutes $\mathbb{F}_{p^2}$, for any $\delta \in\mathbb{F}_{p^2}$ if and only if each prime factor of $p-1$ divides the order of $\frac{1}{2}$ in $\mathbb{F}_{p^2}^*$ but does not divide $p^2-1$.
\end{corollary}
\pf
If $L_1(x)$ is a permutation polynomial over $\mathbb{F}_{p^2}$ such that $L(Tr_{1}^{2}(L_1(x))) =0$,
then from Theorem~\ref{the15} we have $L_1(x) +Tr_{1}^{2}(L_1(x))$ permutes $\mathbb{F}_{p^2}$.
From Proposition~\ref{prop:10} and for any $\delta\in\mathbb{F}_{p^2}$  we have
\[
f(x) = L_1^{-1}( Tr_{1}^{2}(L_1(x))+\delta)+x^p.
\]
and using (\ref{equ2}) we have
\[
\begin{array}{ccl}
f(x) &=& L_1^{-1}(L_1( Tr_{1}^{2}(x))+\delta)+x^p$  for any $x\in\mathbb{F}_{p^2}\\
& =& L_1^{-1}(L_1( Tr_{1}^{2}(x))+ L_1^{-1}(\delta)+x^p\\
& =& Tr_{1}^{2}(x)+ L_1^{-1}(\delta)+x^p\\
&=& x^p + x + L_1^{-1}(\delta)+x^p\\
&=& 2x^p + x + L_1^{-1}(\delta).
\end{array}
\]

To prove that $f$ is injective map, we consider $x$ and  $y \in\mathbb{F}_{p^2}$ satisfy $f(x) = f(y)$.
Then
\[
f(x) = 2x^p + x + L_1^{-1}(\delta) = 2y^p + y + L_1^{-1}(\delta)= f(y),
\]
and hence
\[
2x^p + x  = 2y^p + y,
\]
which is equivalent to

\begin{equation}
\label{eq:4}
( x - y )(2(x - y)^{p-1} - 1) = 0.
\end{equation}

We put $ x - y = z $
then the equation (\ref{eq:4}) becomes
\begin{equation}
\label{eq:8}
z(2z^{p -1} - 1) = 0.
\end{equation}

 By Lemma~\ref{lem3} the polynomial $2z^{p -1} - 1$ is irreducible if and only if each prime factor of $p-1$ divides the order of $\frac{1}{2}$ in $\mathbb{F}_{p^2}^*$ but does not divide $p^2-1$ and $ p \equiv 1 \mod 4$.
  Then the equation (\ref{eq:8}) has a unique solution $ z = 0$.\\
 Finally, we obtain $x = y$ so then $f(x)$ permute $\mathbb{F}_{p^2}.$
\qed

\begin{exe}
Let $ L(x) = x^7+6x$ be a linearized polynomial and $L_1(x) = x^7$ be a linearized permutation polynomial over $\mathbb{F}_{7^2}$, such that $L(Tr_{1}^{2}(L_1(x))) =0$
and $L_1(x)+Tr_{1}^{2}(L_1(x)) = 2x^7 + x$ permute $\mathbb{F}_{7^2}$, then $L_1^{-1}( Tr_{1}^{2}(L_1(x))+\delta)+x^7 = 2x^7 + x + \delta^7$ is a permutation polynomial over $\mathbb{F}_{7^2}$, for any $\delta \in\mathbb{F}_{7^2}$.

\end{exe}

\begin{theorem}
\label{the14}
Let $g(x)$  be a permutation polynomial in $\mathbb{F}_q$ and let $L_{1}(x)$ and $L_{2}(x)\in\mathbb{F}_q$ be two linearized polynomials such that $L_{1}(x)$ is a permutation polynomial and $L_{1}(x)^p = L_{1}(x)$.
If $g(x) + L_{2}(x)$ is also a permutation polynomial, then $g^{-1}(L_2(L_{1}(x))+ \delta)+ L_{1}(x)$ permutes $\mathbb{F}_{q}$.
\end{theorem}
\pf
Let $c$ be an element in $\mathbb{F}_{q}$ which is equivalent to saying that there is $a\in \mathbb{F}_{q}$ such that $c = a^{p}$ with $p$ prime.
Then
\begin{equation}
\label{eq:2}
g^{-1}(L_2(L_{1}(x))+ \delta)+ L_{1}(x) = c,
\end{equation}
which is equivalent to
\[
(L_2(L_{1}(x))+ \delta) = g((a-L_{1}(x))^p).
\]
Let $ y =(a-L_{1}(x))^p  = a^{p} - L_{1}(x)^p $ and $L_{1}(x)= c -y$.
Then $L_{2}(c) + \delta = g(y) + L_{2}(y)$ is a permutation polynomial.
There is a unique $y$ satisfying this equality, so there exists a unique $x \in \mathbb{F}_q $ which satisfies (\ref{eq:2}).
\qed

The following results are easily deduced from Theorem \ref{the14}.
\begin{corollary}
Let $s$ and $t$ be two positive integers.
Consider
 $g(x) = x^s$  and $g^{-1} = x^t$ two permutation polynomials  and let $L_{1}(x)$ be a linearized permutation polynomial in $\mathbb{F}_q$ and let $L_{2}(x)\in\mathbb{F}_q$ be a linearized polynomial with $L_{1}(x)^p = L_{1}(x)$.
 If $g(x) + L_{2}(x)$ is also a permutation polynomial, then $(L_2(L_{1}(x))+ \delta)^t+ L_{1}(x)$ permutes $\mathbb{F}_{q}$, for any $\delta \in \mathbb{F}_{p}$.
\end{corollary}
\begin{corollary}
\label{coro20}
Let $L_{1}$ and $L_{2}(x)$ be a linearized polynomials over $\mathbb{F}_{q}$, such that $L_2$ is a complete linearized polynomial and $L_{1}(x)^p = L_{1}(x)$.
Then $\frac{1}{L_{2}(L_1(x))+ \gamma} + L_{1}(x)$ is a permutation polynomial in $\mathbb{F}_{q}$.
\end{corollary}

\begin{exe}
For $\vartheta$ in $\mathbb{F}_{3^6}^*$  with  $Tr_3 ^6(\vartheta) = 0$, then
 $g(x) = \vartheta^{477}x^{477} $ is a permutation polynomial and $g^{-1}(x) = \vartheta^{-1}x^{29}$ the inverse polynomial of $g(x)$ modulo
$x^{729} - x$.
Moreover,  let $L_{2}(x) =  x $ be a linearized polynomials and $L_{1}(x) = x^{27} + x^9 + x^3 + x $ be a linearized permutation polynomials over $\mathbb{F}_{3^6}$ such that $g(x)+L_{2}(x)$ is a permutation polynomial.\\
 Then $ g^{-1}(L_2(L_{1}(x)))+ L_{1}(x) =
   x + x^3 + x^9 + x^{27} + \vartheta^{-1}(x^3 + 2x^4 + 2x^5 + 2x^7 + 2x^{13} + x^{19} + 2x^{29} + x^{33} + 2x^{39} + x^{45} + 2x^{57} + 2x^{63} + x^{81} + x^{83} + x^{87} + 2x^{91} + 2x^{93} + x^{99} + 2x^{109} + 2x^{111} + 2x^{117} + x^{135} + x^{245} + x^{247} + 2x^{249} + 2x^{253} + 2x^{255} + x^{261} + 2x^{271} + 2x^{279} + x^{297})$ permute $\mathbb{F}_{3^6}.$
\end{exe}
%%%%%%%%%%%%%%%%%%%%%%%%%%%%%%%%%%%%
\subsection{Permutation Polynomials and o-Polynomials}
Our motivation in studying the o-polynomials is to give a link to the complete permutation polynomials over $\mathbb{F}_{q}$. 
\begin{definition}
 A permutation polynomial $f(x)$ over $\mathbb{F}_{q}$ is said o-polynomial if $f(0)=0$ and $ f_v(x)= \frac{f(x+v)+f(v)}{x}$ is a permutation polynomial for all $v\in\mathbb{F}_{q}$, which satisfies $ f_v(0)= 0$.\\
\end{definition}

On the next result we give another characterization of the o-polynomial.
\begin{proposition}
 \label{eq3}

Let $\mathbb{F}_{q}$ be a finite field, a polynomial $f(x)$ is o-polynomial in $\mathbb{F}_{q}$ if and only if
$cf(x)$ is a complete permutation polynomial over $\mathbb{F}_{q}$,  such that  $f(0)=0$,
for any $c$ in $\mathbb{F}_{q}$.
\end{proposition}
\pf

Case 1: Assume that for all $c\neq0$, we have $cf(x)$ is a complete permutation polynomial and consider $ f_v(x)= c$.\\
\begin{equation}
\label{eq10}
f_v(x)=  \frac{f(x+v)+f(v)}{x} = c
\end{equation}
which is equivalent to
\begin{equation}
\label{eq11}
-c^{-1}f(x+v)+ x = c^{-1}f(v).
\end{equation}
For $y = x+v$, then $-c^{-1}f(y)+ y = v + c^{-1}f(v)$ is a permutation polynomial. There is a unique solution y satisfying this equality, then there exists a unique $x\in\mathbb{F}_{q}$ which satisfies the equation(\ref{eq11}) hence a unique solution to equation(\ref{eq10})\\
Case 2: $c = 0$:\\
Since $f$ is a permutation polynomial  and $x+v$ is a linear, then the composed $f(x+v)$ is a permutation polynomial.

If $ f_v(x) = 0$, then there exists a unique $x\in\mathbb{F}_{q}$ satisfied $f(x+v)= -f(v)$.
\qed

%%%%%%%%%%%%%%%%%%%%%%%%%%%%%%%%%%%%%%%%%%%%%%%%%%%%%%%%%%%%%%%%%%%%%%%%%%%%%%%%%
\begin{theorem}
\label{the20}
Let $L(x)$ be a linearized polynomial over $\mathbb{F}_{q}$ and $\gamma \in \mathbb{F}_{p}^*$.
If $g(x)$ is an o-polynomial in $\mathbb{F}_{q}$,
then
\begin{equation}
\label{main:eq}
 g^{-1}(L(x)) + \gamma Tr_{1}^{n}(L(x)),
\end{equation}
permutes $\mathbb{F}_{q}$.
\end{theorem}
\pf
Assume that $L(x)$ is a linearized polynomial over $\mathbb{F}_{q}$, $\gamma \in \mathbb{F}_{p}^*$ and $g(x)$ is an o-polynomial in $\mathbb{F}_{q}$.
If $g^{-1}(L(x)) = y$, then $L(x)= g(y)$ and we have
\[
g^{-1}(L(x)) + \gamma Tr_{1}^{n}(L(x)) = y + \gamma Tr_{1}^{n}(g(y)) = F(y).
\]
Using Lemma~\ref{lem:1.1} where $\omega$ is a $p-th$ root of unity gives
\[
\begin{array}{ccl}
\sum_{y \in \mathbb{F}_{q}}w^{Tr_{1}^{n}(\lambda F(y))}&=& \sum_{y \in \mathbb{F}_{q}}w^{Tr_{1}^{n}(\lambda (y + \gamma Tr_{1}^{n}(g(y))))}\\
&=&\sum_{y \in \mathbb{F}_{q}}w^{Tr_{1}^{n}(\lambda y + \lambda\gamma Tr_{1}^{n}(g(y)))}\\
&=&\sum_{y \in \mathbb{F}_{q}}w^{Tr_{1}^{n}(\lambda y + Tr_{1}^{n}(\lambda\gamma) g(y))}\\
%&=&\sum_{y \in \mathbb{F}_{q}}w^{Tr_{1}^{n}( y + {\lambda}^{-1}Tr_{1}^{n}(\lambda\gamma) g(y))}\\
&=&\sum_{y \in \mathbb{F}_{q}}w^{Tr_{1}^{n}( y + \alpha g(y))}
%&=&\sum_{y \in \mathbb{F}_{q}}w^{Tr_{1}^{n}( y + c g(y)),
\end{array}
\]
where $\alpha = {\lambda}^{-1}Tr_{1}^{n}({\lambda} {\gamma})$ and from Theorem~\ref{eq3}, g(x) is an o-polynomial in
$\mathbb{F}_{q}$ if and only if ${\alpha}g(y)$ is a CPP.
Then
\[
\sum_{y \in \mathbb{F}_{q}}w^{Tr_{1}^{n}(\lambda F(y))} = 0,
\]
for all $\lambda \in\mathbb{F}_{q}^*$.
\qed

\textbf{Remark}: If the polynomials $L(x)$ and $g(x)$ in the hypothesis of Theorem~\ref{the20} are such that $L(x)$ is not a permutation polynomial and $g(x)$ is not a linearized  polynomial, then the polynomial $g^{-1}(L(x))$ is neither a linearized nor a permutation polynomial. This shows that the construction given in Theorem~\ref{the20} is a answer to the Open Problem 1 asked by Charpin and Kyureghyan \cite{kyureghyan1}.
\begin{exe}
\label{exe9}
Let $g(x)= x^6 + x^4 + x^2$ be an o-polynomial on $\mathbb{F}_{2^7}$,
and $g^{-1}(x)= (x+1)^{106} + 1$ be the inverse polynomial of $g(x)$ and let $L(x)= x^2$ be a linearized polynomial over $\mathbb{F}_{2^7}$
so for all $\gamma \neq 0$, we have
\[g^{-1}(L(x)) + \gamma Tr_{1}^{7}(L(x)) = (x^2+1)^{106} + 1 + \gamma( x^2 + x^4 + x^8 + x^{16} + x^{32} + x^{64}).\]

The polynomial $(x^2+1)^{106} + 1 $ is neither a permutation polynomial nor linearized polynomial.
 \end{exe}

 Using a similar proof to the Theorem~\ref{the20}, we obtain  the following theorem.

\begin{theorem}
\label{theo24}
Let $L(x)$ be a linearized polynomial over $\mathbb{F}_{q}$ and $\gamma \in \mathbb{F}_{q}^*$.
If $g(x)$ is an o-polynomial in $\mathbb{F}_{q}$, then
\[
g^{-1}(L(x^{p^i})) + \gamma Tr_{1}^{n}(L(x^{p^i})),
\]
permutes $\mathbb{F}_{q}$.
\end{theorem}

Now we combine Theorem~\ref{the14} and Theorem~\ref{the20} to obtain a new class of permutation polynomials over $\mathbb{F}_q$.
\begin{theorem}
Let $L(x)$, $L_1(x)$ and $L_{2}(x)$ be three linearized polynomials over $\mathbb{F}_{q}$,  such that $L_1(x)$ is permutation polynomial and $L_{1}(x)^p = L_{1}(x)$.
And Let $g(x)$ and $F(x)= g^{-1}(L(x)) + \gamma Tr_{1}^{n}(L(x))$  be two permutation polynomial in $\mathbb{F}_q$.
If $F(x) + L_{2}(x)$ is also a permutation polynomial then $F^{-1}(L_2(L_{1}(x))+ \delta)+ L_{1}(x)$ permutes $\mathbb{F}_{q}$.
\end{theorem}

There are several authors who worked on the permutation polynomials related with Linearized polynomials. For this, in the following Table we give some results on these polynomials related to their references.
\newpage
\begin{table}[h!]
\begin{center}
\begin{tabular}{|c|c|}
\hline

\textbf{The PP in relation with linearized polynomials(LP)over $\mathbb{F}_{q^n}$} & \textbf{ References}\\

\hline

 $h(Tr_{{q^n}/q}(X))\phi(X) +g(Tr_{{q^n}/q}(X))^q - g(Tr_{{q^n}/q}(X))$. With $h$,  $g\in\mathbb{F}_{q^n}$ and $\phi$ is LP in $\mathbb{F}_{q}$& \cite{akbary2}\\

\hline

$L_1(X) +L_2(X)g(L_3(X))$.  With $L_i$ is LP in $\mathbb{F}_{q}$ and $g\in\mathbb{F}_{q^n}$& \cite{akbary2} \\

\hline

 $bL(X) +H(Tr_{{p^n}/p}(X))(L(X) +\gamma)$ is a PP of $\mathbb{F}_{p^n}$ & \cite{Marcos1}\\

\hline

 $L(X) +L(\gamma)G(F(X))$. With $L$ is LP in $\mathbb{F}_{q^n}$,F, $G\in\mathbb{F}_{q}$ and $\gamma\in\mathbb{F}_{q^n}$& \cite{kyureghyan} \\

\hline
 $L(X) +\gamma G(F(X))$. With $L$ is LP in $\mathbb{F}_{q^n}$,F, $G\in\mathbb{F}_{q}$ and $\gamma\in\mathbb{F}_{q^n}$& \cite{kyureghyan}\\

\hline

$L(X) +\gamma Tr_{{p^n}/p}(H(X))$. With $L$ is LP and $H$ in $\mathbb{F}_{q^n}$  & \cite{kyureghyan1}\\

\hline

$g(B(X))+\sum_{i=1}^r (L_i(X) + \gamma_i)h_i(B(X))$. With B and $L_i$ is LP in $\mathbb{F}_{q}$ and $g\in\mathbb{F}_{q^n}$& \cite{Yuan}\\

\hline

$x (L(Tr(x)) + uTr(x) + ux) + vx$.  With $L$ is LP in $\mathbb{F}_{q}$ and $v\in\mathbb{F}_{q}\backslash\{0, 1\} $ and $u\in\mathbb{F}_{q}$& \cite{Baofeng}\\

\hline

$\sum_{i=1}^r (L_i(X) + \gamma_i)h_i(B(X))$ . With B and $L_i$ is LP in $\mathbb{F}_{q}$& \cite{XIAOER}\\

\hline

$L_1(x)+(L_2(x)+)h(Tr_{\mathbb{F}_{q^n}/\mathbb{F}_{q}} (x))$& \cite{XIAOER}\\

\hline
 $L_1(X) +L_2(\gamma)h(f(X))$ is a PP of $\mathbb{F}_{q^n}$ with $L_i$ is LP in $\mathbb{F}_{q}$& \cite{XIAOER}\\

\hline
$g^{-1}(x^p + \delta) + x^p$ is a PP of $\mathbb{F}_{q}$ & Proposition~\ref{prop:10}\\

\hline
$ (L(x^p)+\delta)^t + x^p$ is a PP of $\mathbb{F}_{q}$& Corollary~\ref{coro14}\\
\hline
$ L_1(x) +Tr(L_1(x))$ is a PP of $\mathbb{F}_{p^2} $ with $L_1$ is LP in $\mathbb{F}_{p^2}$ & Corollary~\ref{coro14}\\
\hline
$ L_1^{-1}( Tr(L_1(x))+\delta)+x^p$ is a PP of  $\mathbb{F}_{p^2} $ with $L_1$ is LP in $\mathbb{F}_{p^2}$ & Corollary~\ref{coro17}\\
\hline
$g^{-1}(L_2(L_{1}(x))+ \delta)+ L_{1}(x)$ is a PP of $\mathbb{F}_{q}$. & Theorem~\ref{the14}\\
\hline
$\frac{1}{L_{2}(L_1(x))+ \gamma} + L_{1}(x)$ with $L_i$ is LP in $\mathbb{F}_{q}$ & Corollary~\ref{coro20}\\
\hline
$g^{-1}(L(x^p)) + \gamma Tr(L(x^p))$ with $L$ is LP in $\mathbb{F}_{q}$ & Theorem~\ref{theo24}\\
\hline
\end {tabular}
\end{center}
\caption{Classification Permutation Polynomials in Relation with Linearized Polynomials (LP)}
\end {table}
%Now we will construct a class of permutation trinomial

\end{document}